\documentstyle[12pt]{article}
\begin{document}

\baselineskip 0.65cm

\begin{center}
{\LARGE Bubbles and Wormholes: Analytic Models}
\end{center}

\vspace*{.5cm}

\centerline {\it Anzhong Wang \footnote{e-mail address: wang@on.br}}

\begin{center}
Department of Astrophysics, Observatorio Nacional, Rua General
Jos\'e Cristino $77, 20921-400$ Rio de Janeiro -- RJ, Brazil
\end{center}

\centerline {\it Patricio S. Letelier \footnote{e-mail address:
letelier@ime.unicamp.br} }

\begin{center}
Department of Applied Mathematics-IMECC, Universidade Estadual de Campinas,
$13081-970$ Campinas -- SP,  Brazil
\end{center}

\vspace*{.3cm}

\begin{abstract}

\baselineskip 0.65cm

The first junction conditions of spherically symmetric bubbles are
solved for some cases, and whereby analytic models to the Einstein
field equations are constructed. The effects of bubbles on the
spacetime structure are studied and it is found that in some cases
bubbles can close the spatial sector of the spacetime and turn it into
a compact one, while in other cases they can give rise to wormholes.
One of the most remarkable features of these wormholes is that they do
not necessarily violate the weak and dominant energy condition even at
the classical level.

\vspace{1.5cm}


\end{abstract}

\newpage

\baselineskip 0.9cm

\noindent {\bf I. INTRODUCTION}

Spherically symmetric thin shells or bubbles have been the focus of
interest since the early days of Einstein's General Relativity, for
example, see [1] and references therein, and studied intensively in the
past decade [2 - 13], mainly because of their notable implications to the
inflationary Universe scenario [14]. Most of the investigations [2 - 11]
have been centered in the dynamics of bubbles by using Israel's method
[1]. The advantage of this method is that the four-dimensional coordinates
can be chosen independently in each side of the bubble. Because of this
advantage the relations between the two coordinate systems have been
frequently ignored [4, 8]. The neglect of these conditions is partially
because of their irrelevance to the study of the dynamics
of bubbles and partially because of the complexity of the problem
concerned.

Moreover, the effects of bubbles on the spacetime properties, specially
the global ones, have been hardly studied. However, it has been shown in
the plane-wall case [15] that the existence of these defects could
dramatically change the spacetime geometry.

In the present paper, we shall stress the above issues by considering some
exact solutions to the Einstein field equations, starting from the first
junction conditions. Whereby we are enabled to study the global structure of
the resulting spacetimes. Specifically, the paper is organized as follows: In
Sec. II, using an algorithm [13], we first find the corresponding first
junction conditions. After solving them, we construct exact solutions which
represent spacetimes of bubbles, and then study the spacetime structure. In
Sec. III, our main results are summarized.

\vspace{2.cm}

\noindent {\bf II. EXACT SOLUTIONS OF BUBBLES AND WORMHOLES }

To study solutions that represent static or non-static bubbles and wormholes,
let us begin with the following solution
\begin{equation}
ds^{2} = F dt^{2} + 2G dtdr - H dr^{2} - R^{2}d\Omega ^{2},
\end{equation}
where $d\Omega ^{2} \equiv d\theta^{2} + \sin^{2}\theta d\phi^{2}$, and
\begin{eqnarray}
F &\equiv& \frac{1}{f}(f^{2}T,_{t} ^{2} - R,_{t} ^{2}), \;\;\;\;\;\;
G \equiv \frac{1}{f}(f^{2}T,_{t}T,_{r} - R,_{t}R,_{r}), \nonumber\\
H &\equiv& \frac{1}{f}(R,_{r} ^{2} - f^{2}T,_{r} ^{2}), \;\;\;\;\;\;
f \equiv 1 - \frac{2m}{R},
\end{eqnarray}
and $T, R$ are functions of $t$ and $r$ only. The coordinates will be
numbered as $\{x^{\mu}\} = \{t, r, \theta, \phi\}, \; (\mu = 0, 1, 2, 3)$
with $ - \infty < t, r < + \infty, 0 \le \theta \le \pi,$ and $0 \le
\phi \le 2\pi.$ Note that, {\em provided} that the functions $T$ and $R$
are well defined in terms of $t$ and $r$, the
above solution is essentially the Schwarzschild solution but written in a
different coordinate system. Actually, in terms of $T$ and $R$, one can see
that it will take the form commonly used. The reason to adopt the $(t, r)$
coordinates will be seen clearly in the following discussions.
Following Ref. 13, we make the {\em ansatz}
\begin{equation}
 T = M(\xi - |\psi|) + N(\xi + |\psi|), \;\;\;
 R = U(\xi - |\psi|) + V(\xi + |\psi|),
\end{equation}
where $M, N, U$ and $V$ are at least $C^{4}$ functions of their
indicated arguments in the sense defined in [16], and $\xi, \psi$ are
smooth functions of $t$ and $r$. Note that the notations used here are
slightly different from the ones used in Ref. 13. It is easy to see
that Eq.(3) is well defined, respectively, in the regions $\Omega^{+}$
and $\Omega^{-}$, where $\Omega^{+} \equiv \{x^{\mu}: \psi \ge 0\}$ and
$\Omega^{-} \equiv \{x^{\mu}: \psi \le 0\}$. Therefore, the resulting
solutions in $\Omega^{+}$ and $\Omega^{-}$ are locally isometric to the
Schwarzschild solution. However, across the hypersurface $\psi = 0$ the
functions $T$ and $R$ are only $C^{0}$ with respect to $t$ and $r$. As
a result, the metric coefficients are only $C^{-1}$. It is these
``pathological" coordinate transformations that will lead to new
solutions. As a matter of fact, Eq.(3) is not simply coordinate
transformations. Technically, it is equivalent to cut and then glue two
identical parts of the Schwarzschild solutions together along the
hypersurface $ \psi = 0$. Of cause, such a ``surgery" does not always
give us physically meaningful solutions unless some additional
conditions are imposed. One of our purposes in the following is to find
out these conditions.

Before proceeding, let us first note that the gluing of two
Schwarzschild solutions through a bubble has been considered by several
authors, see, for instance, Refs. 7 and 9. However, our solutions are
different from theirs at least in two points: First, the first junction
conditions are worked out in our case. As a result, the global
structure of the spacetime can be studied easily. Second,  our bubbles
do not necessarily satisfy the ``baratropic" equation of state as do in
Refs. 7, 9 and others.

Inserting Eq.(3) into Eq.(2), we find that all the metric coefficients can
be written in the form
\begin{equation}
 Y = Y^{+} H(\psi) + Y^{-}[1 - H(\psi)],
\end{equation}
where $H(\psi)$ is the Heaviside function, which is one for $\psi \ge 0$
and zero for $\psi < 0$, and $Y^{\pm}$ are the quantities defined in
$\Omega^{\pm}$ respectively. Specifically, we have
\begin{eqnarray}
F^{+} = && \frac{1}{f^{+}}\{ [f^{+ 2}(\dot{M}_{+} + \dot{N}_{-})^{2}
              - (\dot{U}_{+} + \dot{V}_{-})^{2}]\xi,_{t} ^{2} \nonumber\\
        && - 2[f^{+ 2}(\dot{M}^{2}_{+} - \dot{N}^{2}_{-}) -
           (\dot{U}^{2}_{+} - \dot{V}^{2}_{-})]\xi,_{t}\psi,_{t} \nonumber\\
        && + [f^{+ 2}(\dot{M}_{+} - \dot{N}_{-})^{2} - (\dot{U}_{+} -
             \dot{V}_{-})^{2}]\psi,_{t} ^{2}\}, \nonumber\\
G^{+} = && \frac{1}{f^{+}}\{ [f^{+ 2}(\dot{M}_{+} + \dot{N}_{-})^{2}
              - (\dot{U}_{+} + \dot{V}_{-})^{2}]\xi,_{t} \xi,_{r} \nonumber\\
        && - [f^{+ 2}(\dot{M}^{2}_{+} - \dot{N}^{2}_{-}) -
           (\dot{U}^{2}_{+} - \dot{V}^{2}_{-})]
           (\xi,_{t}\psi,_{r} + \xi,_{r}\psi,_{t}) \nonumber\\
        && + [f^{+ 2}(\dot{M}_{+} - \dot{N}_{-})^{2} - (\dot{U}_{+} -
             \dot{V}_{-})^{2}]\psi,_{t} \psi,_{r} \}, \nonumber\\
H^{+} = && \frac{1}{f^{+}}\{ [(\dot{U}_{+} + \dot{V}_{-})^{2} -
            f^{+ 2}(\dot{M}_{+} + \dot{N}_{-})^{2}]\xi,_{r} ^{2}
            \nonumber\\
        && - 2[(\dot{U}^{2}_{+} - \dot{V}^{2}_{-}) -
             f^{+ 2}(\dot{M}^{2}_{+} -
             \dot{N}^{2}_{-})]\xi,_{r}\psi,_{r} \nonumber\\
        && + [ (\dot{U}_{+} - \dot{V}_{-})^{2} - f^{+ 2}(\dot{M}_{+} -
             \dot{N}_{-})^{2}]\psi,_{r} ^{2}\}, \nonumber\\
 R^{+} = && U_{+} + V_{-},
\end{eqnarray}
and the quantities $F^{-}, G^{-}, H^{-}$ and $R^{-}$ can be obtained from the
above equations by the following replacements
\begin{equation}
  M_{+} \rightarrow M_{-}, \;\; N_{-} \rightarrow N_{+}, \;\;
  U_{+} \rightarrow U_{-}, \;\; V_{-} \rightarrow V_{+}, \;\;
  f_{+} \rightarrow f_{-}, \;\; \psi \rightarrow - \psi,
\end{equation}
where
\begin{equation}
  M_{\pm} \equiv M(\xi \mp \psi),\;\;\; f^{\pm} \equiv 1 - \frac{2m}{R^{\pm}}.
\end{equation}
An overdot denotes the ordinary differentiation with respect to the
indicated argument.

Note that the metric coefficients given by Eqs.(5) and (6) are all
continuous across the hypersurface $\psi = 0$, except for the ones
that are proportional to $\psi,_{t}$ or $\psi,_{r}$
that change signs when crossing $\psi = 0$. Therefore, to
have the first junction conditions hold, that is, the metric coefficients
must be at least $C^{0}$ [17], these terms need vanish on the surface,
i.e.,
\begin{equation}
  \dot{M}^{2} - \dot{N}^{2} = \frac{\dot{U}^{2} - \dot{V}^{2}}{f^{2}},
      \;\;\;\;  (\psi = 0).
\end{equation}
On the other hand, since we are concerned with physical bubbles, we require
that the hypersurface $ \psi = 0$ be time-like, $\psi,_{\mu}\psi,_{\nu}
g^{\mu\nu} > 0$, which now reads
\begin{equation}
  E(t, r) \equiv F \psi,_{r} ^{2} - 2 G \psi,_{r}\psi,_{t} -
                   H \psi,_{t} ^{2} > 0, \;\;\; (\psi = 0).
\end{equation}

{}From the results obtained in Ref. 12, we find that corresponding to the
solutions of Eqs. (5) - (9) the energy-momentum tensor (EMT) is given by
\begin{equation}
  T_{\mu \nu} = \tau_{\mu \nu} \delta(\psi),
\end{equation}
where $\delta(\psi)$ is the Dirac delta function, and $\tau_{\mu \nu}$
the surface EMT of the bubble located on the hypersurface $\psi  = 0$,
and given by \footnote{ The surface EMT of the wall is usually defined
as [1-11] $ S_{\mu \nu} \equiv \int{T_{\mu \nu}} dn,$ where $n$ is the
proper distance in the direction perpendicular to the wall.  When $\xi
= t, \psi = \psi(r)$, we have $dn = [H^{1/2}/\psi,_{r}]d\psi$.  Then,
we find $S_{\mu \nu} = [H^{1/2}/\psi,_{r}]|_{\psi = 0}\tau_{\mu \nu}.$
Thus, the surface energy density and tensions of the wall defined here
are different from the ones in [1-11] by a factor
$[H^{1/2}/\psi,_{r}]|_{\psi = 0}.$ }
\begin{equation}
  \tau_{\mu \nu} = \sigma u_{\mu} u_{\nu} -
  \tau (\theta_{\mu}\theta_{\nu} + \phi_{\mu}\phi_{\nu}),
\end{equation}
with
\begin{eqnarray}
 \sigma &=& - \frac{2 E}{FH + G^{2}} \; \frac{[R,_{\psi}]^{-}}
              {R_{0}(\xi)}, \nonumber\\
 \tau  &=& - \frac{1}{2(FH + G^{2})} [ 2E \;
             \frac{[R,_{\psi}]^{-}}{R_{0}(\xi)}
               \nonumber\\
       & & + (\psi,_{r} ^{2}[F,_{\psi}]^{-} - 2 \psi,_{t} \psi,_{r}
             [G,_{\psi}]^{-} - \psi,_{t} ^{2}[H,_{\psi}]^{-}) ],
\end{eqnarray}
\begin{eqnarray}
u_{\mu} &=&  E^{-1/2}[ (F\psi,_{r} - G\psi,_{t})\delta^{t}_{\mu}
           + (G\psi,_{r} + H\psi,_{t})\delta^{r}_{\mu} ], \nonumber\\
\theta_{\mu} &=& R_{0}(\xi)\delta^{\theta}_{\mu}, \;\;\;\;\;\;\;
\phi_{\mu} = \sin\theta R_{0}(\xi)\delta^{\phi}_{\mu}, \nonumber\\
u_{\lambda}u^{\lambda} &=& - \; \theta_{\lambda} \theta^{\lambda} = -
       \; \phi_{\lambda} \phi^{\lambda} = 1,
\end{eqnarray}
and
\begin{equation}
  R_{0}(\xi) \equiv R(\xi,\psi= 0) = U(\xi) + V(\xi).
\end{equation}
{}From Eqs.(11) and (13) we can see that the quantity $\sigma$ represents
the surface energy density of the bubble and $\tau$ its tensions in the
tangential directions\footnote{ See Footnote $3.$}.

To study the above solutions as a whole is too complicated. Thus,
in the following we shall restrict ourselves to the cases where
\begin{equation}
  \xi = t, \;\;\;\;\; \psi = \psi(r).
\end{equation}
Now we need to solve the restraint equation (8).  Since the functions
$M, N, U$ and $V$ are arbitrary, we can always first choose any
functions for three of them and then integrate Eq.(8) to get the other.
This freedom is partially due to  the arbitrary choice of the
coordinates and partially due to the fact that different choice of
these functions implies different matching of the two parts of the
spacetime in each side of the bubble. In the following, we shall
consider the cases with
\begin{equation}
  \dot{M}^{2} - \dot{N}^{2} = \frac{\dot{U}^{2} - \dot{V}^{2}}{f^{2}}= \mu,
\;\;\;\; (\psi = 0),
\end{equation}
where $\mu$ is an arbitrary real constant and must not be confused with
the tensor index. Since for $\mu = 0$ and $\mu \not= 0$ we will have
physically different solutions, we shall consider them separately.

\vspace{.6cm}

\centerline{\bf A. Solutions with $\mu = 0$ }

When $\mu = 0$, Eq.(16) has the solution
\begin{equation}
  M = \epsilon_{1} N + M_{0}, \;\;\;
  U = \epsilon_{2} V + R_{0}, \;\; (\epsilon_{1, 2} = \pm 1),
\end{equation}
where $M_{0}$ and $R_{0}$ are two integration constants, and $N$ and $V$
are arbitrary functions. It can be shown that in the present case only
does the choice $\epsilon_{1} = - \epsilon_{2} = 1$ give physically
meaningful solutions. For this choice, Eq.(12) becomes
\begin{equation}
  \sigma  = \gamma^{-1} \tau = -  \frac{2 (R_{0} - 2m)}{R_{0}^{2} \dot{V}(t)},
  \;\;\;\; \gamma \equiv \frac{ R_{0} - m }{2(R_{0} - 2m)}.
\end{equation}
That is, in the present case the bubbles satisfy the ``baratropic" equation
of state. Thus, they must belong to the solutions studied in Refs. 7 and 9.
On the other hand, Eq.(9) now is equivalent to
\begin{equation}
    R_{0} > 2m,
\end{equation}
which means that the bubbles must be greater than the Schwarzschild sphere.
Combining Eqs.(18) and (19) we find
\begin{equation}
    \gamma = \frac{1}{2} + \frac{m}{2(R_{0} - 2m)} > \frac{1}{2}.
\end{equation}
Thus, in the present case all the bubbles are gravitationally repulsive [7].

To study further this class of solutions, let us consider the cases with
\begin{equation}
    V = - \alpha t^{\beta}, \;\;\; \psi = \prod^{n}_{k=1}{(r - a_{k})},
\end{equation}
where $a_{k} \ge 0, \alpha, \beta $ are constants. Then, we find that
\begin{equation}
    \sigma \delta(\psi) = \gamma^{-1} \tau \delta(\psi) =
    \sum^{n}_{k=1}{\sigma_{k}\delta(r - a_{k})},
\end{equation}
where
\begin{equation}
    \sigma = \frac{\sigma_{0}}{2\alpha \beta}t^{1-\beta}, \;\;\;\;
    \sigma_{k} = \sigma\left\{\prod^{n}_{i\not=k}{|a_{i} -
a_{k}|}\right\}^{-1},
\end{equation}
and $\sigma_{0} \equiv 4(2\gamma - 1)/[m(4\gamma - 1)^{2}]$.
Therefore, the corresponding solutions actually represent $n$ bubbles
that connect $(n+1)$ regions, each of which is locally isometric to
the Schwarzschild solution. Since on each of the bubbles, we have
$R = R_{0}$, all the bubbles have the same physical radius.

{\bf {Case}} $\alpha$): $\alpha = \frac{1}{2}, \beta = 1, a_{1} = R_{0}$,
and $n = 1.$ In this case, we find that the metric is given by
\begin{equation}
ds^{2} = \left\{ \begin{array}{c}
(1 - \frac{2m}{r}) dt^{2} - (1 - \frac{2m}{r})^{-1} dr^{2} - r^{2}d\Omega^{2},
\;\; r \le R_{0}, \\
              \\
(1 - \frac{2m}{2R_{0} - r}) dt^{2} - (1 - \frac{2m}{2R_{0} - r})^{-1} dr^{2}
- (2R_{0} - r)^{2}d\Omega^{2}, \;\; r \ge R_{0}, \end{array} \right.
\end{equation}
and that Eqs.(22) and (23) simply yield $\sigma = \sigma_{0}$.
Therefore, now the solution represents a spherical static bubble with
constant surface energy density and tensions. The solution has two
horizons at $r = 2m, \; 2(R_{0} - m)$ and is singular respectively at $
r = 0, \;2R_{0}$, which seal off the spacetime and turn it into a
compact one (A different interpretation is given in [18]). While its
cosmological interest is not clear, the above solution does show that
the existence of bubbles can dramatically change the spacetime
geometry.

{\bf {Case}} $\beta$): $\alpha = \frac{1}{2}, \beta = 1, a_{2} > a_{1}$,
and $n = 2.$
Then, the solution represents two bubbles that divide the spacetime into
three regions, in each of which the metric takes the form
\begin{equation}
ds^{2} = (1 - \frac{2m}{R}) dt^{2} - 4(r - r_{m})^{2}(1 - \frac{2m}{R})^{-1}
         dr^{2} - R^{2}d\Omega^{2},
\end{equation}
where
\begin{equation}
R = \left\{ \begin{array}{c}
R_{0} - (a_{1} - r)(a_{2} - r),\;\;\; r \le a_{1}, \\
     \\
R_{0} - (r - a_{1})(a_{2} - r),\;\;\; a_{1} \le r \le a_{2}, \\
     \\
R_{0} - (r - a_{1})(r - a_{2}),\;\;\; r \ge a_{2}, \end{array} \right.
\end{equation}
and $r_{m} \equiv (a_{1} + a_{2})/2$. The corresponding surface energy
densities are given by
\begin{equation}
\sigma_{1} = \sigma_{2}  = \frac{\sigma_{0}}{ a_{2} - a_{1} }.
\end{equation}
It can be shown that in general the spacetime is singular at
\begin{equation}
r_{1,2} \equiv r_{m} \pm \sqrt{(a_{2} - a_{1})^{2} + 4R_{0}},\;\;
r_{3,4} \equiv r_{m} \pm \sqrt{(a_{2} - a_{1})^{2} - 4R_{0}},
\end{equation}
and has horizons at
\begin{equation}
r_{5,6} \equiv r_{m} \pm \sqrt{(a_{2} - a_{1})^{2} + 4(R_{0} - 2m)},\;\;
r_{7,8} \equiv r_{m} \pm \sqrt{(a_{2} - a_{1})^{2} - 4(R_{0} - 2m)}.
\end{equation}

Note that in general the maximum number of singular points or event horizons
is $n(n+1)$, where $n$ is the number of bubbles.

{\bf {Case}} $\gamma$): $\alpha = - \frac{1}{2}, \beta = 1, a_{1} = R_{0}$,
and $n = 1.$ Then, we find that
\begin{equation}
ds^{2} = \left\{ \begin{array}{c}
(1 - \frac{2m}{2R_{0} - r}) dt^{2} - (1 - \frac{2m}{2R_{0} - r})^{-1} dr^{2}
- (2R_{0} - r)^{2}d\Omega^{2}, \;\; r \le R_{0},\\
   \\
(1 - \frac{2m}{r}) dt^{2} - (1 - \frac{2m}{r})^{-1} dr^{2} - r^{2}d\Omega^{2},
\;\; r \ge R_{0}. \end{array} \right.
\end{equation}
Comparing Eq.(24) with the above equation we find that these two cases
are related each other by exchanging the form of the metric inside and
outside of the bubble. Because of this exchanging, one can show that in
the present case the spacetime has no singularities and horizons. Also,
as $r \rightarrow \pm \infty$, the spacetime becomes asymptotically
flat. Thus, a remote observer who moves along a time-like radial
geodesic toward the bubble will pass through it within finite time and
soon finds himself in another asymptotically flat region.  Therefore,
in the present case, the bubble acts like the throat of a wormhole
[19]. As shown by Morris and Thorne in [19], the price to construct
such a static wormhole is to violate the weak energy condition (WEC).
In the present model this particular feature is manifested by the fact
that the surface energy density of the bubble is negative, $\sigma = -
\sigma_{0} < 0$.  It should be noted that the above solution was first
studied in [20] in a different manner.

{\bf {Case}} $\delta$): $\alpha = - \frac{1}{2}, \beta = 1, a_{2} >
a_{1}$, and $n = 2.$ Clearly, this corresponds to Case $\beta)$, in
which there are two bubbles with radii $a_{1}$ and $a_{2}$,
respectively. The metric takes the same form as Eq.(25) but with
\begin{equation}
R = \left\{ \begin{array}{c}
R_{0} + (a_{1} - r)(a_{2} - r),\;\;\; r \le a_{1}, \\
     \\
R_{0} + (r - a_{1})(a_{2} - r),\;\;\; a_{1} \le r \le a_{2}, \\
     \\
R_{0} + (r - a_{1})(r - a_{2}),\;\;\; r \ge a_{2}. \end{array} \right.
\end{equation}
The above equation together with Eq.(19) show that the whole spacetime
is free of any singularities and horizons. The region between the two
bubbles is isometric to a compact region of the Schwarzschild
spacetime, and the ones in outside of the two bubbles are
asymptotically flat. Thus, this solution also represents a wormhole
with a finite thickness of throat, $\triangle l =  a_{2} - a_{1},$ and
the ``exotic" matter is concentrated at the two mouths of the throat, $
r = a_{1}$ and $ r = a_{2}$, with
\begin{equation}
\sigma_{1} = \sigma_{2}  = - \frac{\sigma_{0}}{ a_{2} - a_{1} }.
\end{equation}

\vspace{2.cm}

\centerline { {\bf B. Solutions with } $\mu \not= 0$}

Part of the work to be presented in this subsection has been briefly
reported in [21, 22]. In the following, we shall provide a systematic
study.  When $\mu \not= 0$, let us first assume

\begin{equation}
 M = AN + B, \;\;\;\;\;\; U = a V + b,
\end{equation}
where $A, B, a$ and $b$ are arbitrary constants. Then, the integration
of Eq. (16) yields
\begin{eqnarray}
&& N(t) = \varepsilon_1 \mu_1 t + N_0 , \nonumber \\
&& \varepsilon_2 \mu_2 t = V(t) + \frac{2m}{1+a} \ln [R_{0}(t) - 2m]
+ V_0,
\end{eqnarray}
with $N_0$ and $V_0$ being integration constants,
\begin{equation}
\mu_1 \equiv \left(\frac{\mu}{A^2 - 1}\right)^{1/2} ,\;\;
\mu_2 \equiv \left(\frac{\mu}{a^2 - 1}\right)^{1/2} , \;\; \varepsilon_1,
\varepsilon_2 = \pm 1,
\end{equation}
and
\begin{equation}
R_{0}(t) \equiv R(t, \psi=0) = (1 + a) V(t) + b.
\end{equation}
Once Eq. (4) is solved, the metric coefficients of  (2) are in
turn fixed in terms of $t$ and $\psi$. Inserting Eqs.(33) and (34) into
Eq.(12) we find
\begin{equation}
\sigma = \frac{\sigma_0}{R_{0}(t)} , \quad \tau =
\frac{\sigma_0 [R_{0}(t)-m]}{2R_{0}(t) [R_{0}(t) - 2m]},
\end{equation}
where
\begin{equation}
\sigma_0 \equiv \frac{2\varepsilon_2 (1+A)}{ \mu_2 (a - A) }
\end{equation}
is a constant. In the following, we shall choose the free parameters
such that $\sigma_0 > 0$. On the other hand, Eq.(9) now becomes
\begin{equation}
\mu (a - A)(1 + a)(1 + A) > 0.
\end{equation}
Thus, provided Eq.(39) holds, the above solutions represent
a spherical bubble connecting two regions, each of which is locally
isometric to part of the Schwarzschild spacetime. Since the radius of
the bubble $R_{0}(t)$ is time-dependent, the bubble in the present case is
not static. On the other hand, Eqs.(33) and (34) imply that
\begin{equation}
R_{0}(t) > 2m.
\end{equation}
That is, the bubble is  always greater than the Schwarzschild sphere.

{}From Eq.(37) it is easy to show that the bubbles in this
case do not satisfy the ``baratropic" equation of state. Thus, they do
not fall into the solutions studied in Refs. 7 and 9. From the same equation,
we also find that
\begin{equation}
\sigma - \tau = \frac{\sigma_0 [R_{0}(t) - 3m]}{2[R_{0}(t) - 2m]} \ge 0,
\;\;\;\; for \;\; R_{0}(t) \ge  3m.
\end{equation}
Therefore, as long as the radius of the bubble is greater than
or equal to $3m$, it
will satisfy both of the weak and dominant energy conditions [16], although
not the strong one, since for any $R_{0}(t)$ we always have
\begin{equation}
\sigma - 2 \tau =  - \frac{\sigma_0 m}{R_{0}(t) - 2m} < 0.
\end{equation}
That is, the ``Newtonian" mass of the bubble is negative, and the bubble
is always gravitational repulsive [7].

The dynamics of the bubble can be studied using the kinematical
quantities
\begin{equation}
\frac{dR_{0}(t)}{ds} = \beta \left[\frac{R_{0}(t) -
                    2m}{R_{0}(t)}\right]^{1/2}, \;\;\;\;
\frac{d^{2}R_{0}(t)}{ds^{2}} =  \frac{ m \beta^2 }{R_{0}(t)} ,
\end{equation}
where $s$ denotes the proper time measured by observers who are
at rest relative to the bubble, and
\begin{equation}
\beta \equiv \varepsilon_2 \mu_2 (1+a) \left[ \frac{(1 - a)(1 - A)}
{ 2 \mu (a - A) }\right]^{1/2}.
\end{equation}
The above equations show that in the present case the bubble either
expands $(dR_{0}(t)/ds > 0)$ or collapses $(dR_{0}(t)/ds < 0)$, depending
on the choice of the free parameters. From Eqs.(5) and (34), we also have
\begin{equation}
\left.\frac{\partial R(t, |\psi|)}{\partial |\psi|}\right|_{\psi = 0} =
\varepsilon_2 (1 - a)\mu_2 f(R_{0}).
\end{equation}
That is, the topology of the spacetime in the {\em neighborhood} of the
bubble depends on the choice of the parameters $\varepsilon_2$ and
$a$.

In parallel to the last subsection, now let us turn to consider the
following representative cases:

(a) $\varepsilon_2 = -1,\;\; 0 < a < 1, \;\; 0 < A < 1,\;\; A > a,\;\; b = 2m,
\;\; V_0 = 0$, and $\psi = r$. Then, we find
\begin{eqnarray}
Exp (-\mu_2 t) &=& [(1+a)V(t)]^{\frac{2m}{1+a}} Exp[V(t)] , \nonumber\\
R(t, |r|) &=& a V (t-|r|) + V(t+|r|) + 2m, \nonumber\\
R_{0}(t) &=& (1 + a) V(t) + 2m.
\end{eqnarray}
{}From the above equation, it is easy to show that
\begin{equation}
R(t, |r|) = \left\{ \begin{array}{c}
 + \infty, \;\;\; as \; r \rightarrow \pm \infty \;\; at\; a\; moment\; t =
t_{1}, \\
\\
 \ge 2m, \;\;\; for \; any\; t \; and \; r.\;\;\;\;\;\;\;\;\;\;\;\;\;\;\;\;
\;\;\;\;\;\;
 \end{array} \right.
\end{equation}
where $R(t, |r|) = 2m$ only when $t = +\infty$. Thus, in the
present case the bubble acts as the throat of a wormhole that connects
two asymptotically flat Schwarzschild universes. However, this wormhole
is distinguishable from all the known ones in the sense: First, it
satisfies both of the weak and dominant energy conditions. Note that
dynamic wormholes that satisfy the WEC have been recently studied in
[23, 24] in the framework of Einstein theory, and in [25] in the
framework of Brans-Dicke theory. Second, the physical radial coordinate
$R$ is initially decreasing when away from the bubble, as we can see
from Eq.(45), which yields $[\partial R(t, |r|)/\partial |r|] |_{r = 0}
= -(1-a)\mu_{2} f < 0.$ But, as it decreases to a minimum, say,
$R_{min.}$, which is always greater than or equal to $2m$, it starts to
increase.  And as $|r| \rightarrow + \infty$, we have $R(t, |r|)
\rightarrow + \infty.$ Dynamic wormholes made of two asymptotically
flat Schwarzschild spacetimes were also studied by Visser in [20].
Assuming that $R$ is always an increasing function of $|r|$, i.e.,
$[\partial R(t, |r|)/\partial |r|] |_{r = 0} > 0$ for any $r$, Visser
found that all the wormholes necessarily violate the WEC.

It should be noted that in most of the previous studies of bubbles, the
spacetime topology was classified by the signs of the angular
component, $K^{\theta}_{\theta}$, of the extrinsic curvature tensor of
the bubble, where $K^{\theta}_{\theta}$ is given by [26]
\begin{equation}
K^{\theta}_{\theta} = \frac{1}{R_{0}(t)} \left. \frac{\partial R}
{\partial N} \right|_{r = 0},
\end{equation}
$N$ denotes the Gaussian normal coordinate to the bubble. Clearly, this
is correct only for the static case. When the spacetime is
time-dependent, the situation is different. It determines the spacetime
topology only in the {\em neighborhood} of the bubble, and the global
topology of the spacetime could be quite different from the local one.
This fact has been noticed quite recently in [26] and the above
solutions provide another example.  Regarding to the latter, one can
see that all the statements concerning the global structure of the wall
spacetimes given in the previous literature should be taken with some
cautions.

On the other hand, from Eq.(46) we find that
\begin{equation}
R_{0}(t) = \left\{ \begin{array}{c}
+ \infty, \;\;\; as \; t \rightarrow - \infty, \\
\\
2m, \;\;\; as \; t \rightarrow + \infty. \end{array} \right.
\end{equation}
That is, the corresponding solutions represent a collapsing wormhole.
The wormhole throat starts to collapse at $R_{0}(- \infty) = \infty$
and ends at $R_{0}(+ \infty) = 2m$. On the other hand, from Eq. (43) we
find that
\begin{equation}
 \triangle{s} = \frac{1}{|\beta|} \int^{\infty}_{2m}\frac{x dx}{x - 2m}
            = \infty.
\end{equation}
Thus, to complete the process of collapse, the throat will take an
infinitely long proper time. Consequently, a space adventurer will have
enough time to pass through the throat of the wormhole from one
asymptotically flat region to the other before the radius of the throat
shrinks to $2m$, where  the event horizon usually appearing in the
Schwarzschild solution is developed.

To further study the above solutions, let us consider the embedding of
them in the  three dimensional Euclidean space
\begin{equation}
dl^2 = dZ^2 + dR^2 + R^2 d^2 \phi.
\end{equation}
Because of the reflection symmetry of the spacetime, it is sufficient
to consider the problem only in the region where $r \ge 0$. Then,
following Ref. [19], we find that
\begin{equation}
\left(\frac{dZ}{dr}\right)^2 = (a \dot{V}_{+} - \dot{V}_{-})^{-2}\{
(a \dot{V}_{+} - \dot{V}_{-})^{2}[1 - f(R)] -
\mu_{1} ^{2}(1 - A)^{2}f(R)\}.
\end{equation}
One can show that the right side of the above equation changes signs
from point to point. That is, in the present case our solutions can not
be embedded in a 3-dimensional Euclidean space and pictured as an
ordinary Euclidean curved surface.  Recall that not any two dimensional
metric can be embedded into a three dimensional Euclidean space.
Classical examples are the Moebius strip and the
Gauss-B\'olyai-Lobachevski metric $ds^2 = (1+r^2)^{-1} dr^2 + r^2
d\phi^2$ [27].

(b) $\varepsilon_2 = -1,\;\; 0 < a < 1, \;\; 0 < A < 1,\;\; A > a,\;\;
b = 2m, \;\; V_0 = 0$, and $\psi = (r - a_{2})(r - a_{1})$, where $
a_{2} > a_{1} > 0$.  Clearly, in this case we have two bubbles, located
respectively on the hypersurfaces $r = a_{1}$ and $r = a_{2}$. The
physical radii of these two bubbles are the same and are give by Eqs.
(34) and (36). Each of the two bubbles collapses in the same way as the
single bubble given in Case (a). In particular, they collapse from
$R_{0}(- \infty) = \infty$ to $R_{0}(+ \infty) = 2m$ by taking an
infinitely long proper time. The two bubbles connect three regions: $ -
\infty < r < a_{1}, a_{1} < r < a_{2}$, and $ a_{2} < r < + \infty$,
each of which is locally isometric to the Schwarzschild spacetime. In
the region in between the two bubbles ($a_{1} < r < a_{2}$), the
spacetime is compact with the physical radius always being greater than
or equal to $2m$, where equality holds only when $t \rightarrow +
\infty$. The two regions outside of the bubbles are reflection
symmetric with respect to the hypersurface $ r = (a_{2} + a_{1})/2$.
They are asymptotically flat and free of any singularities and horizons
at any finite time. Therefore, they also represent wormholes that
satisfy the weak and dominant energy conditions. The only difference
between solutions given in this case and the ones given in the last is
that the thickness of the throat of the wormhole now is different from
zero and equal to $\triangle l = a_{2} - a_{1}$, and that the two
bubbles act as two mouths of the throat.

(c) $\varepsilon_2 = 1, \;\; a > 1,\;\; A > 1,\;\; a > A, \;\; b = 2m,
V_0 = 0$, and $ \psi = r$. In this case, we have
\begin{eqnarray}
Exp (\mu_2 t) &=& [(1+a)V(t)]^{\frac{2m}{1+a}} Exp[V(t)] , \nonumber\\
R(t, |r|) &=& a V (t-|r|) + V(t+|r|) + 2m, \nonumber\\
R_{0}(t) &=& (1 + a) V(t) + 2m.
\end{eqnarray}
Similar to Case (a), one can show that these solutions
represent an expanding bubble, which connects two asymptotically flat
universes and satisfy the weak and dominant energy conditions as long
as $R_{0}(t) \ge 3m$. The bubble expands from $R_{0}(- \infty) = 2m$ to
$R_{0}(+ \infty) = + \infty$ by taking an infinite long proper time.

{}From Eq. (37) we can see that when $R_{0}(t)$ is approaching $2m$, the
tensions in the tangent directions of the bubble tend to infinite.
Thus, in the course of the collapse of the bubble, as described in
Case (a), it is not difficult to imagine that the bubble will
explode due to the enormous tensions before its radius really shrinks to
$2m$. By properly arranging the parameters, the explosion could
happen before the trapped surface is developed. After the explosion,
the material may recompose and form another wormhole, the later
evolution of which follows more or less the same process as described
by the solutions given in the present case.

(d) $\varepsilon_2 = 1, \;\; a > 1,\;\; A > 1,\;\; a > A, \;\; b = 2m,
V_0 = 0$, and $\psi = (r - a_{2})(r - a_{1})$, where $ a_{2} > a_{1} > 0$.
Clearly, this is the time-reversed procces of Case (b), and corresponds
two expanding bubbles. All the properties of this class of solutions can
be obtained by the replacement $t$ by $- t$ in the solutions of
Case (b).

\vspace{2.cm}

\noindent{\bf III. CONCLUSIONS AND DISCUSSIONS}

In this paper, by solving the first junction conditions, we have
constructed some analytic solutions to the Einstein field equations,
which represent multiple bubbles connecting regions that are locally
isometric to the Schwarzschild solution. The obtained solutions can be
classified into two categories, one represents static bubbles and the
other dynamic bubbles.  For the static bubbles, provided that their
surface energy densities are positive, the spacetimes are always
compact and closed by curvature singularities. For the dynamic bubbles,
the resulted spacetimes represent wormholes. However, these wormholes
are distinguishable from the known ones in the sense that they satisfy
both the weak and dominant energy conditions, but violate the strong
one. Therefore, in contrast to the static ones [19], dynamic wormholes
can be built without violating the WEC.  Note that the violation of the
strong energy condition nowadays does not seem to be a very serious
drawback. Recall that cosmic bubbles and domain walls formed in the
early Universe do not satisfy this condition either.

The recent studies of wormholes usually fall into two different
directions.  One is concerned with the energy conditions [28], and the
other is concerned with the vacuum polarization due to the quantum
effects [29 - 31].  To the first, one can see that even it can be shown
that the WEC is preserved at the quantum level for the generic cases,
the existence of wormholes can not be ruled out. As shown in this
paper, they can exist even in the classical level without violating the
WEC. To the second, Hawking [29, 30] argued that, when the vacuum
polarization effects are taken into account,  one might finally show
that such a building of a traversable wormhole is impossible, although
Thorne and others [31] seem to defend the opposite opinion.  The
considerations of the latter now are under investigation, and the
results will be discussed somewhere else.

\vspace{2.cm}

\noindent {\bf{ACKNOLEDGMENTS}}

Part of the work was done when one of the authors (A.W.) was visiting
the Department of Applied Mathematics, UNICAMP. He thanks the
Department for the hospitality. The financial assistance  from FAPESP
and CNPq is gratefully acknowledged.

\end{document}